\documentclass[preprint,double-spaced,noshowpacs,preprintnumbers,amsmath,amssymb]{revtex4-2}
\usepackage{graphicx}% Include figure files
\usepackage{dcolumn}% Align table columns on decimal point
\usepackage{bm,slashed}% bold math
\PassOptionsToPackage{hyphens}{url}
\usepackage[colorlinks=true]{hyperref}
\newcommand{\hodge}{{\star}}
\begin{document}

\preprint{}

\title{The Dirac equation as a linear tensor equation for one component}% Force line breaks with \\

\author{Andrey Akhmeteli}
 \email{akhmeteli@ltasolid.com}
 \homepage{http://www.akhmeteli.org}
\affiliation{%
LTASolid Inc.\\
10616 Meadowglen Ln 2708\\
Houston, TX 77042, USA}

%\affiliation{

\date{\today}% It is always \today, today,

\begin{abstract}
The Dirac equation is one of the most fundamental equations of modern physics. It is a spinor equation, but some tensor equivalents of the equation were proposed previously. Those equivalents were either nonlinear or involved several components of the Dirac field. On the other hand, the author showed previously that the Dirac equation in electromagnetic field is equivalent to a fourth-order equation for one component of the Dirac spinor. The equivalency is used in this work to derive a linear tensor equivalent of the Dirac equation for just one component. This surprising result can be used in applications of the Dirac equation, for example,  in general relativity or for lattice approximation of the Dirac field and can improve our understanding of the Dirac equation.
\end{abstract}

\maketitle

\section{\label{sec:level1a}Introduction}
``Of all the equations of physics, perhaps the most `magical' is the Dirac equation. It is the most freely invented, the least conditioned by experiment, the one with the strangest and most startling consequences''~\cite{Wilmag}. The Dirac equation is a spinor equation, but previously, some authors offered tensor equations equivalent to the Dirac equation. Those equations were either nonlinear ~\cite{Whit} or, in the case of the Dirac-K\"{a}hler equation (~\cite{Darwin28},~\cite{Iwa},~\cite{Kahl}), had a large number of components. The approach of ~\cite{March} allows to significantly reduce the number of components, but still requires several components.

The surprising result of this work is that it is possible to derive a linear tensor equation for just one component that is generally equivalent to the Dirac equation. This result builds on the previous work ~\cite{Akhmeteli-JMP,Akhm2015,Akhmeteliqr} (see also ~\cite{Bagrov2014}, pp. 24-25, ~\cite{Bagro}), where it was shown that, in a general case, three out of four complex components of the Dirac spinor can be algebraically eliminated from the Dirac equation in an arbitrary electromagnetic field. Therefore, the Dirac equation is generally equivalent to a fourth-order partial differential equation for just one component, which can be made real (at least locally) by a gauge transformation.

One usually emphasizes the advantages of tensor, rather than spinor, equations for applications in general relativity ~\cite{Whit} and lattice approximation of Dirac fields ~\cite{Becher}, but there can be other applications of the fundamental result of this work in the future. As Feynman said in his Nobel lecture, "a good theoretical physicist today might find it useful to have a wide range of physical viewpoints and mathematical expressions of the same theory"~\cite{Feynnob}.

Section \ref{sec:level1a0} summarizes the results of the prevous work ~\cite{Akhmeteli-JMP,Akhm2015,Akhmeteliqr}  used in the present article.

Section \ref{sec:level1a1}  contains the general derivation of the linear tensor equation for one component that is equivalent to the Dirac equation.

Formulations of the  linear tensor equation in terms of antisymmetric second-rank tensors and 3D vectors are provided, respectively, in Sections \ref{sec:level1n2} and \ref{sec:level1n3}.

The ancillary files include a Mathematica notebook ~\cite{Akhmnb} and its .pdf version  ~\cite{Akhmpdf}. The notebook contains the proofs of several algebraic statements (usually indicated by phrases ``One can check that\ldots'') in Sections \ref{sec:level1a1}, \ref{sec:level1n2}, and \ref{sec:level1n3}, using the chiral representation of gamma-matrices, which is sufficient for our purpose. Proofs of the statements in Section \ref{sec:level1a0} can be found in  ~\cite{Akhm2015,Akhmeteliqr}.
\section{\label{sec:level1a0}The Dirac equation as a fourth-order linear equation for one component}
To make this article reasonably self-contained, let us provide some notation and results from ~\cite{Akhm2015,Akhmeteliqr}.

We use the Dirac equation in the form:
\begin{equation}\label{eq:pr25}
(i\slashed{\partial}-\slashed{A})\psi=\psi,
\end{equation}
where, e.g., $\slashed{A}=A_\mu\gamma^\mu$ (the Feynman slash notation). A system of units $\hbar=c=m=1$ is used, and the electric charge $e$ is included in $A_\mu$ ($eA_\mu \rightarrow A_\mu$). The metric tensor is
\begin{equation}\label{eq:metric}
g=\left(g^{\mu\nu}\right)=
\left( \begin{array}{cccc}
1 & 0 & 0& 0\\
0 & -1 & 0 & 0\\
0 & 0 &-1 & 0\\
0 &0 & 0 & -1
 \end{array} \right).
\end{equation}
Multiplying both sides of (\ref{eq:pr25}) by $(i\slashed{\partial}-\slashed{A})$ from the left and using notation
\begin{equation}\label{eq:li2}
\sigma^{\mu\nu}=\frac{i}{2}[\gamma^\mu,\gamma^\nu],
\end{equation}
\begin{eqnarray}\label{eq:d8n1}
F^{\mu\nu}=A^{\nu,\mu}-A^{\mu,\nu}=\left( \begin{array}{cccc}
0 & -E^1 & -E^2 & -E^3\\
E^1 & 0 & -H^3 & H^2\\
E^2 & H^3 & 0 & -H^1\\
E^3 &-H^2 & H^1 & 0  \end{array} \right),
\end{eqnarray}
\begin{equation}\label{eq:li4}
F=\frac{1}{2}F_{\nu\mu}\sigma^{\nu\mu},
\end{equation}
we obtain:
\begin{equation}\label{eq:li3}
(\Box'+F)\psi=0,
\end{equation}
where the modified d'Alembertian $\Box'$ is defined as follows:
\begin{eqnarray}\label{eq:d8n2}
\Box'=\partial^\mu\partial_\mu+2 i A^\mu\partial_\mu+i A^\mu_{,\mu}-A^\mu A_\mu+1=-(i\partial_\mu-A_\mu)(i\partial^\mu-A^\mu)+1.
\end{eqnarray}
We assume that the set of $\gamma$-matrices satisfies the standard hermiticity conditions ~\cite{Itzykson}:
\begin{equation}\label{eq:li5a}
\gamma^{\mu\dag}=\gamma^0\gamma^\mu\gamma^0, \gamma^{5\dag}=\gamma^5=i\gamma^0\gamma^1\gamma^2\gamma^3.
\end{equation}
Then a charge conjugation  matrix $C$ can be chosen in such a way (~\cite{Bogol},~\cite{Schweber}) that
\begin{equation}\label{eq:li5}
C\gamma^\mu C^{-1}=-\gamma^{\mu T}, C\gamma^5 C^{-1}=\gamma^{5 T}, C\sigma^{\mu\nu}C^{-1}=-\sigma^{\mu\nu T},
\end{equation}
\begin{equation}\label{eq:li6}
C^T=C^\dag=-C, CC^\dag=C^\dag C=I, C^2=-I,
\end{equation}
where the superscript $T$ denotes transposition, and $I$ is the unit matrix. The properties of the charge conjugation matrix (\ref{eq:li5},\ref{eq:li6}) are extensively used in the following.

We choose constant Dirac spinors $\xi$ and $\eta$ (they do not depend on the spacetime coordinates $x=(x^0,x^1,x^2,x^3)$, so, e. g., $\partial_\mu\xi\equiv 0$) in such a way that they are both eigenvectors of $\gamma^5$ with the same eigenvalue $+1$ or $-1$, and the normalization condition
\begin{equation}\label{eq:norm7}
\bar{\xi}\eta^c=1
\end{equation}
is satisfied, where $\eta^c$ is the charge conjugated spinor
\begin{equation}\label{eq:lin14eta}
\eta^c=C\bar{\eta}^T.
\end{equation}
Due to the normalization condition, the spinors  $\xi$ and $\eta$ are linearly independent. If they are eigenvectors of $\gamma^5$ with the same eigenvalue $\pm 1$, the Dirac conjugated spinors $\bar{\xi}$ and $\bar{\eta}$ are linearly independent left eigenvectors of $\gamma^5$ with the same eigenvalue $\mp 1$. One can check that spinors $\bar{\xi}F$ and $\bar{\eta}F$ are also left eigenvectors of $\gamma^5$ with the same eigenvalue $\mp 1$, so, as the space of left eigenvectors of $\gamma^5$ with eigenvalue $\mp 1$ is two-dimensional, $\bar{\xi}F$ and $\bar{\eta}F$ are linear combinations of $\bar{\xi}$ and $\bar{\eta}$. One can calculate the coefficients and obtain:
\begin{equation}\label{eq:lin12mod}
\bar{\xi}F=(\bar{\xi}F\eta^c)\bar{\xi}-(\bar{\xi}F\xi^c)\bar{\eta}=f_{\xi\eta}\bar{\xi}-f_{\xi\xi}\bar{\eta},
\end{equation}
\begin{equation}\label{eq:lin13mod}
\bar{\eta}F=(\bar{\eta}F\eta^c)\bar{\xi}-(\bar{\eta}F\xi^c)\bar{\eta}=f_{\eta\eta}\bar{\xi}-f_{\eta\xi}\bar{\eta},
\end{equation}
where the following notation  for components of the electromagnetic field is used:
\begin{equation}\label{eq:simp}
f_{\alpha \beta}=\bar{\alpha}F\beta^c.
\end{equation}
Here $\alpha$ and $\beta$ are some Dirac spinors. 

Multiplying (\ref{eq:li3}) by $\bar{\xi}$ and $\bar{\eta}$ from the left and using (\ref{eq:lin12mod}) and (\ref{eq:lin13mod}), we obtain:
\begin{eqnarray}\label{eq:lin21}
\nonumber
\Box'(\bar{\xi}\psi)+f_{\xi\eta}(\bar{\xi}\psi)-f_{\xi\xi}(\bar{\eta}\psi)=0,
\\
\nonumber
\Box'(\bar{\eta}\psi)+f_{\eta\eta}(\bar{\xi}\psi)-f_{\eta\xi}(\bar{\eta}\psi)=0,
\end{eqnarray}
so
\begin{equation}\label{eq:lin22}
\bar{\eta}\psi=f_{\xi\xi}^{-1}(\Box'(\bar{\xi}\psi)+f_{\xi\eta}(\bar{\xi}\psi))
\end{equation}
and
\begin{equation}\label{eq:sim2}
((\Box'-f_{\xi\eta})f_{\xi\xi}^{-1}(\Box'
+f_{\xi\eta})+f_{\eta\eta})(\bar{\xi}\psi)=0
\end{equation}
(one can show that $f_{\xi\eta}=f_{\eta\xi}$). Thus, it is possible to algebraically eliminate three out of four components of the Dirac equation from (\ref{eq:li3}) and derive an equation for the remaining component $\bar{\xi}\psi$.

One can derive the recipes for calculation of the other components of the Dirac spinor. If $\bar{\xi}\psi$ is known, another component, $\bar{\eta}\psi$, can be determined using equation (\ref{eq:lin22}). Then $\psi$ can be expressed as a sum of a right-handed and a left-handed spinors $\psi_+$ and $\psi_-$, where $\gamma^5 \psi_\pm=\pm\psi_\pm$:
\begin{eqnarray}\label{eq:lin26}
\psi=\psi_++\psi_-,
\\
\psi_\pm=\frac{1}{2}(1\pm\gamma^5)\psi.
\end{eqnarray}
Then one can check that
\begin{equation}\label{eq:lin31}
\psi_\mp=(\bar{\xi}\psi)\eta^c-(\bar{\eta}\psi)\xi^c.
\end{equation}
When $\psi_\mp$ is found in this way, $\psi_{\pm}$ can be found using the Dirac equation (\ref{eq:pr25}):
\begin{equation}\label{eq:lin32}
(i\slashed{\partial}-\slashed{A})\psi_\mp=\frac{1}{2}(i\slashed{\partial}-\slashed{A})(1\mp\gamma^5)\psi=
\frac{1}{2}(1\pm\gamma^5)(i\slashed{\partial}-\slashed{A})\psi=\psi_\pm,
\end{equation}
thus, the Dirac spinor can be fully restored if the component $\bar{\xi}\psi$ is known. Equation (\ref{eq:sim2}),  together with the recipes for calculation of the other components of the Dirac spinor, is generally equivalent to the Dirac equation. Some "non-transversal" cases where there is no equivalency, for example, if $f_{\xi\xi}\equiv 0$, do not seem to be important from the experimental point of view. For example, $f_{\xi\xi}\equiv 0$ for a free Dirac particle, and it is an important case theoretically, but an arbitrarily weak electromagnetic field, for example, the Coulomb field of a single charged particle in the Universe, restores the equivalency.

The denominator $f_{\xi\xi}$ in (\ref{eq:sim2}) may look inaesthetic. One can get rid of it by multiplying (\ref{eq:sim2}) by $(f_{\xi\xi})^3$, but this is a matter of preference.
\section{\label{sec:level1a1}Derivation of the linear tensor equivalent of the Dirac equation}

Let us consider a Dirac spinor $\chi$. It can be presented as a sum of two chiral spinors $\chi_+$ and $\chi_-$ that are eigenvectors of $\gamma^5$ with eigenvalues of $+1$ and $-1$, respectively:

\begin{equation}\label{eq:chi1}
\chi=\chi_++\chi_-, \gamma^5\chi_+=\chi_+, \gamma^5\chi_-=-\chi_-.
\end{equation}
In the chiral representation of $\gamma$-matrices ~\cite{Itzykson},
\begin{equation}\label{eq:d1}
\gamma^0=\left( \begin{array}{cc}
0 & -I\\
-I & 0 \end{array} \right),\gamma^i=\left( \begin{array}{cc}
0 & \sigma^i \\
-\sigma^i & 0 \end{array} \right),\gamma^5=\left( \begin{array}{cc}
I & 0\\
0 & -I \end{array} \right),C=\left( \begin{array}{cc}
-i\sigma^2 & 0\\
0 & i\sigma^2 \end{array} \right),
\end{equation}
where index $i$ runs from 1 to 3, and $\sigma^i$ are the Pauli matrices. Then $\chi$, $\chi_+$, and $\chi_-$ have the following components:
\begin{equation}\label{eq:chi3}
\chi=\left( \begin{array}{c}
\chi_1\\
\chi_2\\
\chi_3\\
\chi_4\end{array}\right),
\chi_+=\left( \begin{array}{c}
\chi_1\\
\chi_2\\
0\\
0\end{array}\right),
\chi_-=\left( \begin{array}{c}
0\\
0\\
\chi_3\\
\chi_4\end{array}\right).
\end{equation}
Similarly, let us introduce spinors $\zeta$, $\zeta_+$, and $\zeta_-$ and consider the following matrices $\theta_+$ and $\theta_-$:
\begin{eqnarray}\label{eq:chi3a}
\nonumber
\left(\theta_+^{\mu\nu}\right) =
\left(\overline{\chi_+}  \sigma^{\mu\nu}\zeta_+^c\right)=
\\
\left( \begin{array}{cccc}
0 & i(\chi_1^*\zeta_1^*-\chi_2^*\zeta_2^*) & \chi_1^*\zeta_1^* +\chi_2^*\zeta_2^*&  -i(\chi_2^*\zeta_1^*+\chi_1^*\zeta_2^*) \\
-i(\chi_1^*\zeta_1^*-\chi_2^*\zeta_2^*) & 0 &\chi_2^*\zeta_1^*+\chi_1^*\zeta_2^* & -i(\chi_1^*\zeta_1^* +\chi_2^*\zeta_2^*)\\
 -\chi_1^*\zeta_1^* -\chi_2^*\zeta_2^* & -\chi_2^*\zeta_1^*-\chi_1^*\zeta_2^* & 0 &-\chi_1^*\zeta_1^* +\chi_2^*\zeta_2^*\\
i(\chi_2^*\zeta_1^*+\chi_1^*\zeta_2^*) & i(\chi_1^*\zeta_1^* +\chi_2^*\zeta_2^*)& \chi_1^*\zeta_1^* -\chi_2^*\zeta_2^* & 0  \end{array} \right)
\end{eqnarray}
and
\begin{eqnarray}\label{eq:chi3b}
\nonumber
\left(\theta_-^{\mu\nu}\right) =
\left(\overline{\chi_-}  \sigma^{\mu\nu}\zeta_-^c\right)=
\\
\left( \begin{array}{cccc}
0 & i(\chi_3^*\zeta_3^*-\chi_4^*\zeta_4^*) & \chi_3^*\zeta_3^* +\chi_4^*\zeta_4^*&  -i(\chi_4^*\zeta_3^*+\chi_3^*\zeta_4^*) \\
-i(\chi_3^*\zeta_3^*-\chi_4^*\zeta_4^*) & 0 &-\chi_4^*\zeta_3^*-\chi_3^*\zeta_4^* & i(\chi_3^*\zeta_3^* +\chi_4^*\zeta_4^*)\\
 -\chi_3^*\zeta_3^* -\chi_4^*\zeta_4^* & \chi_4^*\zeta_3^*+\chi_3^*\zeta_4^* & 0 &\chi_3^*\zeta_3^* -\chi_4^*\zeta_4^*\\
i(\chi_4^*\zeta_3^*+\chi_3^*\zeta_4^*) &- i(\chi_3^*\zeta_3^* +\chi_4^*\zeta_4^*)& -\chi_3^*\zeta_3^* +\chi_4^*\zeta_4^* & 0  \end{array} \right).
\end{eqnarray}
Proper ortochronous Lorentz transformations acting on Dirac spinors in the chiral representation have the following form~\cite{Macf}:
\begin{equation}\label{eq:chi3c}
\left( \begin{array}{cc}
s &  \\
 & (s^\dagger)^{-1}
 \end{array} \right),
\end{equation}
where
\begin{equation}\label{eq:chi3d}
s=\left( \begin{array}{cc}
s_{11} & s_{12}  \\
 s_{21} &s_{22}
 \end{array} \right),
(s^\dagger)^{-1}=\left( \begin{array}{cc}
s_{22}^* & -s_{21}^*  \\
- s_{12}^* &s_{11}^*
 \end{array} \right),
\end{equation}
and the determinant $|s|=1$. The relevant Lorentz transformation acting on tensors is
\begin{equation}\label{eq:chi3e}
\Lambda=\left(\Lambda^{\mu}_{\ \nu}\right)=
\frac{1}{2}\left( \begin{array}{cccc}
s_{11} & s_{12} & s_{21} & s_{22}\\
s_{21} & s_{22} & s_{11} & s_{12}\\
-i s_{21} & -i s_{22} & i s_{11} & i s_{12}\\
s_{11} & s_{12} & -s_{21} & -s_{22}
 \end{array} \right)
\left( \begin{array}{cccc}
s_{11}^* & s_{12}^* & -i s_{12}^* & s_{11}^*\\
s_{12}^* & s_{11}^* & i s_{11}^* &- s_{12}^*\\
s_{21}^* & s_{22}^* & -i s_{22}^* & s_{21}^*\\
s_{22}^* & s_{21}^* & i s_{21}^* &-s_{22}^*
\end{array} \right).
\end{equation}
The derivation of  (\ref{eq:chi3e}) is similar to that in~\cite{Macf}. Our (\ref{eq:chi3}) differs from equation (62) of ~\cite{Macf}, ultimately, because of a different choice of the sign of $\gamma^0$. One can check that $g_{\mu\nu}\Lambda^{\mu}_{\ \rho}\Lambda^{\nu}_{\ \lambda}=g_{\rho\lambda}$~\cite{Macf} , or $\Lambda^T g_l \Lambda=g_l$, where the metric tensor with lower indices $g_l=(g_{\mu\nu})$, and that replacement of $\chi_1,\chi_2,\zeta_1,\zeta_2$ in $\theta_+$ with $\chi_1^{\prime},\chi_2^{\prime},\zeta_1^{\prime},\zeta_2^{\prime}$ in accordance with formulas
\begin{eqnarray}\label{eq:chi3f}
\left(\begin{array}{c}
\chi_1^{\prime}\\
\chi_2^{\prime}\end{array}\right)=
s
\left(\begin{array}{c}
\chi_1\\
\chi_2\end{array}\right),
\left(\begin{array}{c}
\zeta_1^{\prime}\\
\zeta_2^{\prime}\end{array}\right)=
s\left(\begin{array}{c}
\zeta_1\\
\zeta_2\end{array}\right)
\end{eqnarray}
gives the same result as $\Lambda^{\mu}_{\ \rho}\Lambda^{\nu}_{\ \lambda}\theta_+^{\rho\lambda}$, or $\Lambda \theta_+\Lambda^T$, so $\theta_+$ transforms as an antisymmetric second-rank tensor. Also, replacement of $\chi_3,\chi_4,\zeta_3,\zeta_4$ in $\theta_-$ with $\chi_3^{\prime},\chi_4^{\prime},\zeta_3^{\prime},\zeta_4^{\prime}$ in accordance with formulas
\begin{eqnarray}\label{eq:chi3fsecond}
\left(\begin{array}{c}
\chi_3^{\prime}\\
\chi_4^{\prime}\end{array}\right)=
(s^\dagger)^{-1}
\left(\begin{array}{c}
\chi_3\\
\chi_4\end{array}\right),
\left(\begin{array}{c}
\zeta_3^{\prime}\\
\zeta_4^{\prime}\end{array}\right)=
(s^\dagger)^{-1}\left(\begin{array}{c}
\zeta_1\\
\zeta_2\end{array}\right)
\end{eqnarray}
gives the same result as $\Lambda^{\mu}_{\ \rho}\Lambda^{\nu}_{\ \lambda}\theta_-^{\rho\lambda}$, or $\Lambda \theta_-\Lambda^T$, so $\theta_-$ transforms as an antisymmetric second-rank tensor. Similarly, $\theta_\pm^*$ (complex conjugates of $\theta_\pm$) also transform as  antisymmetric second-rank tensors.

One can check that 
\begin{equation}\label{eq:dual1}
\left(\theta^{\mu\nu}_\pm\right)=\left(\mp i\hodge\theta^{\mu\nu}_\pm\right),\left(\theta^{*\mu\nu}_\pm\right)=\left(\pm i\hodge\theta^{*\mu\nu}_\pm\right),
\end{equation}
where the Hodge dual of a second-rank antisymmetric tensor is defined as ~\cite{Jackson}
\begin{equation}\label{eq:dual1b}
\hodge F^{\alpha\beta}=\frac{1}{2}\epsilon^{\alpha\beta\gamma\delta} F_{\gamma\delta},
\end{equation}
and $\epsilon^{\alpha\beta\gamma\delta}$ is the totally antisymmetric Levi-Civita tensor ($\epsilon^{0 1 2 3}=1$). That means that tensors $\theta^{\mu\nu}_\pm$ are fully defined by 3D vectors $\theta^{0 i}_\pm$, and vice versa, similar to how the Weber vector (Riemann-Silberstein vector) $\bm{E}+i \bm{H}$ ~\cite{Kies} fully defines the real antisymmetric second-rank electromagnetic field tensor $F^{\mu\nu}$. Four-dimensional rotations in the 4D Minkowski space for the tensors correspond to rotations through complex angles in a 3D complex space for the 3D vectors (~\cite{Landau}, \S 25). The following could be presented in terms of 3D vectors, but the author will mostly use tensors and add a short section on 3D vectors. No distinction is drawn between tensors and pseudotensors in this work.
 
Chiral spinors $\psi_\pm$ can be represented by antisymmetric second-rank tensors  (~\cite{Whit},~\cite{Cartan}, Section 154):
\begin{equation}\label{eq:chi1second}
\left(\psi_\pm^{\mu\nu}\right) =\left(\overline{\psi_\pm^c}  \sigma^{\mu\nu}\psi_\pm\right) =\left(\psi_\pm^T C \sigma^{\mu\nu}\psi_\pm\right) .
\end{equation}.
One can see that these are tensors as $\left(\psi_\pm^{\mu\nu}\right)$ coincide with $\left(\theta_\mp^{\mu\nu}\right)$ if $\chi=\psi^c$ and $\zeta=\psi^c$ (one can check that $(\psi^c)^c=\psi$ and $\overline{\psi^c}=\psi^TC$).
One can check that
\begin{equation}\label{eq:dual3}
\psi_\pm^{\mu\nu}\psi_{\pm\mu\nu}=0.
\end{equation} 

So is it possible to replace the spinor elements of (\ref{eq:sim2}) by tensor ones while preserving linearity? Note that we want to avoid the following: while chiral spinors can be represented by antisymmetric second-rank tensors satisfying (\ref{eq:dual1},\ref{eq:dual3}), such tensors define spinors only up to a sign, so we need to avoid the ambiguity related to square roots.

It seems natural to start with the tensors corresponding to the spinors $\xi^c$ and $\eta^c$, however such an approach creates the above-mentioned ambiguity. Therefore, we start with a tensor
\begin{equation}\label{eq:dual15}
u^{\mu\nu}=(\xi^c)^T C \sigma^{\mu\nu}\xi^c=(C\bar{\xi}^{\,T})^T C\sigma^{\mu\nu}\xi^c=\bar{\xi} C^T C\sigma^{\mu\nu}\xi^c=\bar{\xi}\sigma^{\mu\nu}\xi^c,
\end{equation}
corresponding to $\xi^c$, and a tensor $v^{\mu\nu}=(\xi^c)^T C \sigma^{\mu\nu}\eta^c=\bar{\xi}\sigma^{\mu\nu}\eta^c$. These tensors satisfy (\ref{eq:dual1}), where the signs are chosen depending on whether $\xi$ and $\eta$ are eigenvectors of $\gamma^5$ with eigenvalue $+1$ or $-1$. Taking into account the normalization condition  (\ref{eq:norm7}), one can show for the tensor $w^{\mu\nu}=(\eta^c)^T C \sigma^{\mu\nu}\eta^c=\bar{\eta}\sigma^{\mu\nu}\eta^c$ corresponding to the chiral spinor $\eta^c$ that
\begin{equation}\label{eq:dual16}
u^{\mu\nu}w_{\mu\nu}=-8, v^{\mu\nu}w_{\mu\nu}=0, w^{\mu\nu}w_{\mu\nu}=0.
\end{equation}
We also have
\begin{equation}\label{eq:dual17}
u^{\mu\nu}u_{\mu\nu}=0, u^{\mu\nu}v_{\mu\nu}=0,v^{\mu\nu}v_{\mu\nu}=4.
\end{equation}
There is only one solution for $w^{\mu\nu}$ satisfying (\ref{eq:dual16}) and  (\ref{eq:dual1}) if $u^{\mu\nu}$ and $v^{\mu\nu}$ satisfy (\ref{eq:dual17}) and $u^{\mu\nu}\neq 0$. To find it, one can first choose any tensor $k^{\mu\nu}$ (satisfying (\ref{eq:dual1})) such that $k^{\mu\nu}k_{\mu\nu}=0$ and $u^{\mu\nu}k_{\mu\nu}\neq 0$. Let us prove it.

First, let us consider an antisymmetric second-rank tensor $a^{\mu\nu}$ satisfying (\ref{eq:dual1}), so
\begin{equation}\label{eq:dual18a}
\left(a^{\mu\nu}\right)=\left( \begin{array}{cccc}
0 & a^{01} & a^{02} & a^{03}\\
-a^{01} & 0 & \pm i a^{03} & \mp i a^{02}\\
-a^{02} &\mp i a^{03} & 0 & \pm i a^{01}\\
-a^{03} &\pm i a^{02} & \mp i a^{01} &0
 \end{array} \right).
\end{equation}
Such tensor is fully defined by the 3D vector $\bm{a}=(a^1,a^2,a^3)=(a^{01},a^{02},a^{03})$. Let us define the scalar product of (complex) 3D vectors $\bm{a}$ and $\bm{b}=(b^1,b^2,b^3)$ as $(\bm{a}\cdot\bm{b})=a^i b^i=a^1 b^1+a^2 b^2+a^3 b^3$ (no conjugation). One can check that $a^{\mu\nu}b_{\mu\nu}=-4(\bm{a}\cdot\bm{b})$, where $a^{\mu\nu}$ and $b^{\mu\nu}$ are antisymmetric second-rank tensors corresponding to $\bm{a}$ and $\bm{b}$ and satisfying  (\ref{eq:dual1}).

 The space of antisymetric tensors satisfying (\ref{eq:dual1}) is 3-dimensional, so one can choose such an antisymmetric tensor $t^{\mu\nu}$ satisfying (\ref{eq:dual1}) that $u^{\mu\nu}$, $v^{\mu\nu}$, and $t^{\mu\nu}$ are linearly independent.

Let us prove that $u^{\mu\nu}t_{\mu\nu}\neq 0$. Indeed, otherwise $(\bm{u}\cdot\bm{t})=0$. Let us consider the following matrix (similar to the Gram matrix, where, however, a different scalar product is used for complex vectors):
\begin{equation}\label{eq:dual19a}
M=\left( \begin{array}{ccc}
(\bm{u}\cdot\bm{u}) & (\bm{u}\cdot\bm{v}) & (\bm{u}\cdot\bm{t}) \\
(\bm{v}\cdot\bm{u}) & (\bm{v}\cdot\bm{v}) & (\bm{v}\cdot\bm{t}) \\
(\bm{t}\cdot\bm{u}) &(\bm{t}\cdot\bm{v}) & (\bm{t}\cdot\bm{t})\\
 \end{array} \right)=Q^T Q,
\end{equation}
where
\begin{equation}\label{eq:dual20a}
Q=\left( \begin{array}{ccc}
u^1 & v^1 & t^1 \\
u^2 & v^2 & t^2 \\
u^3 & v^3 & t^3\\
 \end{array} \right).
\end{equation}
From (\ref{eq:dual17}), we have $(\bm{u}\cdot\bm{u})=(\bm{u}\cdot\bm{v})=0$, and we assumed $(\bm{u}\cdot\bm{t})=0$, so we have for the determinants:
\begin{equation}\label{eq:dual21a}
0=|M|=|Q|^2,
\end{equation}
So $|Q|=0$, although we assumed that $u^{\mu\nu}$, $v^{\mu\nu}$, and $t^{\mu\nu}$, and, therefore, $\bm{u}$, $\bm{v}$, and $\bm{t}$ are linearly independent. Therefore, the assumption $(\bm{u}\cdot\bm{t})=0$ leads to a contradiction. Thus, we proved that $u^{\mu\nu}t_{\mu\nu}\neq 0$.

Let us seek $k^{\mu\nu}$ in the form $(\alpha u^{\mu\nu}+v^{\mu\nu}+t^{\mu\nu})$. Then we have:
\begin{equation}\label{eq:dual22a}
0=(\alpha u^{\mu\nu}+v^{\mu\nu}+t^{\mu\nu})(\alpha u_{\mu\nu}+v_{\mu\nu}+t_{\mu\nu})=v^{\mu\nu}v_{\mu\nu}+t^{\mu\nu}t_{\mu\nu}+2\alpha u^{\mu\nu}t_{\mu\nu}+2 v^{\mu\nu}t_{\mu\nu},
\end{equation} 
so we have exactly one solution for $\alpha$:
\begin{equation}\label{eq:dual23a}
\alpha=-\frac{v^{\mu\nu}v_{\mu\nu}+t^{\mu\nu}t_{\mu\nu}+2 v^{\mu\nu}t_{\mu\nu}}{2 u^{\mu\nu}t_{\mu\nu}},
\end{equation}
and for $k^{\mu\nu}=\alpha u^{\mu\nu}+v^{\mu\nu}+t^{\mu\nu}$ we have $k^{\mu\nu}k_{\mu\nu}=0$ and  $u^{\mu\nu}k_{\mu\nu}\neq 0$, as otherwise $u^{\mu\nu}t_{\mu\nu}= 0$.

The tensors $u^{\mu\nu},v^{\mu\nu},t^{\mu\nu}$ are linearly independent, so the tensors $u^{\mu\nu},v^{\mu\nu},k^{\mu\nu}$ are also linearly independent, so one can seek the solution  for $w^{\mu\nu}$ in the form
\begin{equation}\label{eq:dual24a}
w^{\mu\nu}=\alpha_1 u^{\mu\nu}+\alpha_2 v^{\mu\nu}+\alpha_3 k^{\mu\nu}.
\end{equation}
Contracting (\ref{eq:dual24a}) with $u_{\mu\nu}$, we obtain
\begin{equation}\label{eq:dual25a}
\alpha_3 u_{\mu\nu}k^{\mu\nu}=-8.
\end{equation}
Contracting (\ref{eq:dual24a}) with $v_{\mu\nu}$, we obtain
\begin{equation}\label{eq:dual26a}
4\alpha_2+\alpha_3 v_{\mu\nu}k^{\mu\nu}=0.
\end{equation}
Contracting (\ref{eq:dual24a}) with $w_{\mu\nu}=\alpha_1 u_{\mu\nu}+\alpha_2 v_{\mu\nu}+\alpha_3 k_{\mu\nu}$, we obtain
\begin{equation}\label{eq:dual27a}
0=(\alpha_2)^2 v_{\mu\nu}v^{\mu\nu}+2\alpha_1\alpha_3 u_{\mu\nu}k^{\mu\nu}+2\alpha_2\alpha_3 v_{\mu\nu}k^{\mu\nu}.
\end{equation}
One can obtain from (\ref{eq:dual25a},\ref{eq:dual26a},\ref{eq:dual27a}):
\begin{equation}\label{eq:dual5}
\alpha_1=-\frac{(v^{\mu\nu}k_{\mu\nu})^2}{(u^{\mu\nu}k_{\mu\nu})^2}, \alpha_2= 2\frac{v^{\mu\nu}k_{\mu\nu}}{u^{\mu\nu}k_{\mu\nu}}, \alpha_3= -\frac{8}{u^{\mu\nu}k_{\mu\nu}}.
\end{equation}

The above is a coordinate-free solution. A more explicit solution (where a specific tensor $k^{\mu\nu}$ is chosen) can be written in coordinates.

If $\xi$ and $\eta$ are eigenvectors of $\gamma^5$ with eigenvalue $\pm 1$ and
\begin{equation}\label{eq:dual6}
u=\left( \begin{array}{cccc}
0 & u_1 & u_2 & u_3\\
-u_1 & 0 & \pm i u_3 & \mp i u_2\\
-u_2 & \mp i u_3 & 0 & \pm i u_1\\
-u_3 & \pm i u_2 & \mp i u_1 & 0
 \end{array} \right),
v=\left( \begin{array}{cccc}
0 & v_1 & v_2 & v_3\\
-v_1 & 0 & \pm i v_3 & \mp i v_2\\
-v_2 & \mp i v_3 & 0 & \pm i v_1\\
-v_3 & \pm i v_2 & \mp i v_1 & 0
 \end{array} \right),
\end{equation}
and
\begin{eqnarray}\label{eq:dual7}
\nonumber
u^{\mu\nu}u_{\mu\nu}=-4((u_1)^2+(u_2)^2+(u_3)^2)=0,
\\
\nonumber
u^{\mu\nu}v_{\mu\nu}=-4(u_1 v_1+u_2 v_2+u_3 v_3)=0,
\\
v^{\mu\nu}v_{\mu\nu}=-4((v_1)^2+(v_2)^2+(v_3)^2)=4,
\end{eqnarray}
tensor $k^{\mu\nu}$ can be chosen in the following form:
\begin{equation}\label{eq:dual8}
k=\left( \begin{array}{cccc}
0 & u_1^* & u_2^* & u_3^*\\
-u_1^* & 0 & \pm i u_3^* & \mp i u_2^*\\
-u_2^* & \mp i u_3^* & 0 & \pm i u_1^*\\
-u_3^* & \pm i u_2^* & \mp i u_1^* & 0
 \end{array} \right),
\end{equation}
then
\begin{equation}\label{eq:dual9}
u^{\mu\nu}k_{\mu\nu}=-4(|u_1|^2+|u_2|^2+|u_3|^2).
\end{equation}
This value does not vanish unless $u=0$.
When tensors $u^{\mu\nu},v^{\mu\nu},w^{\mu\nu}$ are known, it is possible to calculate the components of the electromagnetic field from (\ref{eq:sim2}):
\begin{equation}\label{eq:dual14}
f_{\xi\xi}=\frac{1}{2}F_{\mu\nu}u^{\mu\nu},f_{\xi\eta}=\frac{1}{2}F_{\mu\nu}v^{\mu\nu},f_{\eta\eta}=\frac{1}{2}F_{\mu\nu}w^{\mu\nu}.
\end{equation}
To rewrite (\ref{eq:sim2}) in terms of tensors completely, one needs to find an appropriate form for $\bar{\xi}\psi$.

If $\xi$ is an eigenvector of $\gamma^5$ with eigenvalue $+1$,
\begin{equation}\label{eq:dual15b}
\bar{\xi}\psi=\bar{\xi}\psi_-=-\psi_3 \xi_1^* - \psi_4 \xi_2^*.
\end{equation}
The tensor corresponding to $\psi_-$ is 
\begin{equation}\label{eq:dual16b}
\psi_-^{\mu\nu}=\psi_-^T C \sigma^{\mu\nu}\psi_-.
\end{equation}
One can check that
\begin{equation}\label{eq:dual17b}
\psi_-^{\mu\nu}u_{\mu\nu}=-8(\bar{\xi}\psi)^2.
\end{equation}

If $\xi$ is an eigenvector of $\gamma^5$ with eigenvalue $-1$,
\begin{equation}\label{eq:dual18b}
\bar{\xi}\psi=\bar{\xi}\psi_+=-\psi_1 \xi_3^* - \psi_2 \xi_4^*.
\end{equation}
The tensor corresponding to $\psi_+$ is 
\begin{equation}\label{eq:dual19b}
\psi_+^{\mu\nu}=\psi_+^T C \sigma^{\mu\nu}\psi_+.
\end{equation}
One can check that
\begin{equation}\label{eq:dual20b}
\psi_+^{\mu\nu}u_{\mu\nu}=-8(\bar{\xi}\psi)^2.
\end{equation}
 
Thus, if $\xi$ is an eigenvector of $\gamma^5$ with eigenvalue $\pm 1$,
\begin{equation}\label{eq:dual21b}
\bar{\xi}\psi=\left(-\frac{\psi_\mp^{\mu\nu}u_{\mu\nu}}{8}\right)^{\frac{1}{2}}.
\end{equation}
So $\bar{\xi}\psi$ can be expressed using tensors as a square root of a scalar, but as (\ref{eq:sim2}) is a linear equation with respect to $\bar{\xi}\psi$, this does not create any ambiguities related to the sign of the square root (when the sign of the square root is chosen in one point, the choice can be expanded by continuity, at least locally). Thus,  (\ref{eq:sim2}) can be expressed using tensors only.

One can also show that the Dirac current can be found using tensors only. Up to a constant factor, the current is
\begin{equation}\label{eq:dual22b}
j^{\mu}=\bar{\psi}\gamma^{\mu}\psi=j_+^{\mu}+j_-^{\mu},
\end{equation}
where
\begin{equation}\label{eq:dual23b}
j_{\pm}^{\mu}=\bar{\psi}_{\pm}\gamma^{\mu}\psi_{\pm}.
\end{equation}
\maketitle
One can check that for tensors $\psi_{\pm}^{\mu\nu}$ (defined by (\ref{eq:dual16b}) and (\ref{eq:dual19b})) the following is true:
\begin{equation}\label{eq:dual24b}
j_{\pm}^{\mu\nu}=g_{\sigma\lambda}\psi_{\pm}^{\sigma\mu}\left(\psi_{\pm}^{\lambda\nu}\right)^*=-2 j_{\pm}^{\mu}j_{\pm}^{\nu}.
\end{equation}
Thus, if we know tensors $\psi_{\pm}^{\mu\nu}$, we know tensors $ j_{\pm}^{\mu}j_{\pm}^{\nu}$, and, if we know the latter tensors, we can find vectors $ j_{\pm}^{\mu}$, as
\begin{equation}\label{eq:dual25b}
j_{\pm}^{\mu}=\frac{j_{\pm}^{\mu}j_{\pm}^0}{\sqrt{(j_{\pm}^0)^2}}.
\end{equation}
The non-negative value of the square root should be chosen in (\ref{eq:dual25b}) as we know that $j_{\pm}^0\geq 0$. It is not clear whether the option to choose the negative value of the square root and therefore obtain negative charge density for the tensor equivalent of the Dirac equation can be advantageous, e. g., to better describe the antiparticles.

Thus, to prove that one can find out $j^{\mu}$ from one of the scalars $\psi_{\mp}^{\mu\nu}u_{\mu\nu}$, it is sufficient to prove that one can find both tensors $\psi_{\pm}^{\mu\nu}$ from the scalar.  If $\xi$ and $\eta$ are eigenvectors of $\gamma^5$ with eigenvalue $\pm 1$, then $\bar{\xi} \psi=\bar{\xi} \psi_{\mp}$. We have (\ref{eq:lin31}), and
\begin{eqnarray}\label{eq:dual27b}
\nonumber
\psi_{\mp}^{\mu\nu}=\psi_{\mp}^T C \sigma^{\mu\nu}\psi_{\mp}=((\bar{\xi}\psi)(\eta^c)^T-(\bar{\eta}\psi)(\xi^c)^T) C \sigma^{\mu\nu}((\bar{\xi}\psi)\eta^c-(\bar{\eta}\psi)\xi^c)=
\\
\nonumber
(\bar{\xi}\psi)^2(\bar{\eta}\sigma^{\mu\nu}\eta^c)-(\bar{\eta}\psi)(\bar{\xi}\psi)(\bar{\xi}\sigma^{\mu\nu}\eta^c)-(\bar{\xi}\psi)(\bar{\eta}\psi)(\bar{\eta}\sigma^{\mu\nu}\xi^c)+(\bar{\eta}\psi)^2(\bar{\xi}\sigma^{\mu\nu}\xi^c)=
\\
(\bar{\xi}\psi)^2 w^{\mu\nu}-2(\bar{\eta}\psi)(\bar{\xi}\psi) v^{\mu\nu}+(\bar{\eta}\psi)^2 u^{\mu\nu},
\end{eqnarray}
as
\begin{eqnarray}\label{eq:dual28b}
\nonumber
\bar{\eta}\sigma^{\mu\nu}\xi^c=(\bar{\eta}\sigma^{\mu\nu}\xi^c)^T=(\xi^c)^T(\sigma^{\mu\nu})^T(\bar{\eta})^T=(C\bar{\xi}^{\,T})^T(-C\sigma^{\mu\nu}C^{-1})(-C)\eta^c=
\\
\bar{\xi} C^T(C\sigma^{\mu\nu}C)(-C)\eta^c=\bar{\xi}\sigma^{\mu\nu}\eta^c.
\end{eqnarray}
Using (\ref{eq:lin22}) and (\ref{eq:dual14}), we obtain
\begin{equation}\label{eq:dual29b}
\bar{\eta}\psi=(F_{\mu\nu}u^{\mu\nu})^{-1}(2\Box^{'}+F_{\mu\nu}v^{\mu\nu})(\bar{\xi}\psi).
\end{equation}
Using (\ref{eq:dual21b}) and (\ref{eq:dual29b}), we can express products $(\bar{\xi}\psi)^2$, $(\bar{\eta}\psi)(\bar{\xi}\psi)$, and $(\bar{\eta}\psi)^2$ in (\ref{eq:dual27b}) via $\psi_{\mp}^{\mu\nu}u_{\mu\nu}$, and the results do not depend on the choice of the value of the square root in (\ref{eq:dual21b}). Therefore, we can express $\psi_{\mp}^{\mu\nu}$ via the scalar function  $\psi_{\mp}^{\mu\nu}u_{\mu\nu}$ (and its derivatives).

Let us now express $\psi_{\pm}^{\mu\nu}$ via the same scalar function. Using (\ref{eq:lin32}) and (\ref{eq:lin31}), we obtain:
\begin{equation}\label{eq:dual30b}
\psi_{\pm}=\gamma^{\mu}(i\partial_{\mu}-A_{\mu})\psi_{\mp}=\gamma^{\mu}(i\partial_{\mu}-A_{\mu})((\bar{\xi}\psi)\eta^c-(\bar{\eta}\psi)\xi^c)=\gamma^{\mu}(B_{\mu}\xi^c+C_{\mu}\eta^c),
\end{equation}
where
\begin{equation}\label{eq:dual31b}
B_{\mu}=-(i\partial_{\mu}-A_{\mu})(\bar{\eta}\psi),C_{\mu}=(i\partial_{\mu}-A_{\mu})(\bar{\xi}\psi).
\end{equation}
We have
\begin{eqnarray}\label{eq:dual32b}
\nonumber
\psi_{\pm}^{\,T}C=(\gamma^{\mu}(B_{\mu}\xi^c+C_{\mu}\eta^c))^{\,T}C=(B_{\mu}(\xi^c)^{\,T}+C_{\mu}(\eta^c)^{\,T})(-C\gamma^{\mu}C^{-1})C=
\\
(B_{\mu}\bar{\xi}C^{\,T}+C_{\mu}\bar{\eta}C^{\,T})(-C\gamma^{\mu})=-(B_{\mu}\bar{\xi}+C_{\mu}\bar{\eta})\gamma^{\mu}.
\end{eqnarray}
Then
\begin{eqnarray}\label{eq:dual33b}
\nonumber
\psi_{\pm}^{\nu\sigma}=\psi_{\pm}^{\,T}C\sigma^{\nu\sigma}\psi_{\pm}=-(B_{\mu}\bar{\xi}+C_{\mu}\bar{\eta})\gamma^{\mu}\sigma^{\nu\sigma}\gamma^{\lambda}(B_{\lambda}\xi^c+C_{\lambda}\eta^c)=
\\
-B_{\mu}B_{\lambda}\bar{\xi}\gamma^{\mu}\sigma^{\nu\sigma}\gamma^{\lambda}\xi^c-C_{\mu}B_{\lambda}\bar{\eta}\gamma^{\mu}\sigma^{\nu\sigma}\gamma^{\lambda}\xi^c-B_{\mu}C_{\lambda}\bar{\xi}\gamma^{\mu}\sigma^{\nu\sigma}\gamma^{\lambda}\eta^c-C_{\mu}C_{\lambda}\bar{\eta}\gamma^{\mu}\sigma^{\nu\sigma}\gamma^{\lambda}\eta^c.
\end{eqnarray}
Let us use the following formula for the product of four $\gamma$-matrices (~\cite{Kuusela}, p. 17):
\begin{eqnarray}\label{eq:dual34b}
\nonumber
\gamma^{\mu}\gamma^{\nu}\gamma^{\sigma}\gamma^{\lambda}=g^{\mu\nu}g^{\sigma\lambda}-g^{\mu\sigma}g^{\nu\lambda}+g^{\mu\lambda}g^{\nu\sigma}+
\\
g^{\mu\nu}\gamma^{\sigma\lambda}-g^{\mu\sigma}\gamma^{\nu\lambda}+g^{\mu\lambda}\gamma^{\nu\sigma}+g^{\nu\sigma}\gamma^{\mu\lambda}-g^{\nu\lambda}\gamma^{\mu\sigma}+g^{\sigma\lambda}\gamma^{\mu\nu}+\gamma^{\mu\nu\sigma\lambda},
\end{eqnarray}
where the antisymmetrized products of $\gamma$-matrices are
\begin{eqnarray}\label{eq:dual35b}
\nonumber
\gamma^{\mu\nu}=\begin{cases}
  \gamma^{\mu}\gamma^{\nu} \text{ if $\mu\neq\nu$,}  
\\  0 \text{ otherwise}    
\end{cases},
\\
\gamma^{\mu\nu\sigma\lambda}=\begin{cases}
  \gamma^{\mu}\gamma^{\nu}\gamma^{\sigma}\gamma^{\lambda} \text{ if $\{\mu,\nu,\sigma,\lambda\}$ is a permutation of \{0,1,2,3\}},  
\\  0 \text{ otherwise}    
\end{cases},
\end{eqnarray}
so, if $\nu\neq\sigma$,
\begin{eqnarray}\label{eq:dual36b}
\nonumber
\gamma^{\mu}\sigma^{\nu\sigma}\gamma^{\lambda}=\frac{i}{2}\gamma^{\mu}(\gamma^{\nu}\gamma^{\sigma}-\gamma^{\sigma}\gamma^{\nu})\gamma^{\lambda}=i(g^{\mu\nu}g^{\sigma\lambda}-g^{\mu\sigma}g^{\nu\lambda}+
\\
g^{\mu\nu}\gamma^{\sigma\lambda}-g^{\mu\sigma}\gamma^{\nu\lambda}+g^{\mu\lambda}\gamma^{\nu\sigma}-g^{\nu\lambda}\gamma^{\mu\sigma}+g^{\sigma\lambda}\gamma^{\mu\nu}+\gamma^{\mu\nu\sigma\lambda}).
\end{eqnarray}
Let us also note that
\begin{eqnarray}\label{eq:dual37b}
\nonumber
\bar{\eta}\gamma^{\mu}\sigma^{\nu\sigma}\gamma^{\lambda}\xi^c=(\bar{\eta}\gamma^{\mu}\sigma^{\nu\sigma}\gamma^{\lambda}\xi^c)^T=\xi^{cT}\gamma^{\lambda T}\sigma^{\nu\sigma T}\gamma^{\mu T}\bar{\eta}^T=
\\
-\bar{\xi} C (-C\gamma^{\lambda}C^{-1})(-C\sigma^{\nu\sigma}C^{-1}) (-C\gamma^{\mu}C^{-1})(-C\eta^c)=\bar{\xi}\gamma^{\lambda}\sigma^{\nu\sigma}\gamma^{\mu}\eta^c,
\end{eqnarray}
so
\begin{equation}\label{eq:dual38b}
-C_{\mu}B_{\lambda}\bar{\eta}\gamma^{\mu}\sigma^{\nu\sigma}\gamma^{\lambda}\xi^c=-C_{\mu}B_{\lambda}\bar{\xi}\gamma^{\lambda}\sigma^{\nu\sigma}\gamma^{\mu}\eta^c=-B_{\mu}C_{\lambda}\bar{\xi}\gamma^{\mu}\sigma^{\nu\sigma}\gamma^{\lambda}\eta^c,
\end{equation}
and (\ref{eq:dual33b}) becomes
\begin{equation}\label{eq:dual39b}
\psi_{\pm}^{\nu\sigma}=
-B_{\mu}B_{\lambda}\bar{\xi}\gamma^{\mu}\sigma^{\nu\sigma}\gamma^{\lambda}\xi^c-2 B_{\mu}C_{\lambda}\bar{\xi}\gamma^{\mu}\sigma^{\nu\sigma}\gamma^{\lambda}\eta^c-C_{\mu}C_{\lambda}\bar{\eta}\gamma^{\mu}\sigma^{\nu\sigma}\gamma^{\lambda}\eta^c.
\end{equation}
Now we need to substitute (\ref{eq:dual36b}) in (\ref{eq:dual39b}). Let us show how some of the terms can be calculated. For example (one can check that $\bar{\xi}\xi^c=\bar{\eta}\eta^c=0$),
\begin{equation}\label{eq:dual40b}
\bar{\xi}g^{\mu\nu}g^{\sigma\lambda}\eta^c=g^{\mu\nu}g^{\sigma\lambda},\bar{\xi}g^{\mu\nu}g^{\sigma\lambda}\xi^c=\bar{\eta}g^{\mu\nu}g^{\sigma\lambda}\eta^c=0,
\end{equation}
\begin{equation}\label{eq:dual41b}
\bar{\xi}g^{\mu\nu}\gamma^{\sigma\lambda}\eta^c=g^{\mu\nu}\frac{1}{i}v^{\sigma\lambda},\bar{\xi}g^{\mu\nu}\gamma^{\sigma\lambda}\xi^c=g^{\mu\nu}\frac{1}{i}u^{\sigma\lambda},\bar{\eta}g^{\mu\nu}\gamma^{\sigma\lambda}\eta^c=g^{\mu\nu}\frac{1}{i}w^{\sigma\lambda},
\end{equation}
\begin{equation}\label{eq:dual42b}
\bar{\xi}\gamma^{\mu\nu\sigma\lambda}\eta^c=\bar{\xi}\epsilon^{\mu\nu\sigma\lambda}\gamma^0\gamma^1\gamma^2\gamma^3\eta^c=\epsilon^{\mu\nu\sigma\lambda}\bar{\xi}\frac{1}{i}\gamma^5\eta^c=\mp\epsilon^{\mu\nu\sigma\lambda}\frac{1}{i},\bar{\xi}\gamma^{\mu\nu\sigma\lambda}\xi^c=\bar{\eta}\gamma^{\mu\nu\sigma\lambda}\eta^c=0.
\end{equation}

Thus. we have, if $\nu\neq\sigma$,
 \begin{eqnarray}\label{eq:dual43b}
\nonumber
\psi_{\pm}^{\nu\sigma}=-B_{\mu}B_{\lambda}(g^{\mu\nu}u^{\sigma\lambda}-g^{\mu\sigma}u^{\nu\lambda}+g^{\mu\lambda}u^{\nu\sigma}-g^{\nu\lambda}u^{\mu\sigma}+g^{\sigma\lambda}u^{\mu\nu})-
\\
\nonumber
2 B_{\mu}C_{\lambda}(i g^{\mu\nu}g^{\sigma\lambda}-i g^{\mu\sigma}g^{\nu\lambda}+g^{\mu\nu}v^{\sigma\lambda}-g^{\mu\sigma}v^{\nu\lambda}+g^{\mu\lambda}v^{\nu\sigma}-g^{\nu\lambda}v^{\mu\sigma}+g^{\sigma\lambda}v^{\mu\nu}\mp \epsilon^{\mu\nu\sigma\lambda})-
\\
\nonumber
C_{\mu}C_{\lambda}(g^{\mu\nu}w^{\sigma\lambda}-g^{\mu\sigma}w^{\nu\lambda}+g^{\mu\lambda}w^{\nu\sigma}-g^{\nu\lambda}w^{\mu\sigma}+g^{\sigma\lambda}w^{\mu\nu})=
\\
\nonumber
-B^{\nu}B_{\lambda}u^{\sigma\lambda}+B^{\sigma}B_{\lambda}u^{\nu\lambda}-B^{\lambda}B_{\lambda}u^{\nu\sigma}+B_{\mu}B^{\nu}u^{\mu\sigma}-B_{\mu}B^{\sigma}u^{\mu\nu}-2 i B^{\nu}C^{\sigma}+2 i B^{\sigma}C^{\nu}-
\\
\nonumber
2 B^{\nu}C_{\lambda}v^{\sigma\lambda}+2 B^{\sigma}C_{\lambda}v^{\nu\lambda}-2 B^{\lambda}C_{\lambda}v^{\nu\sigma}+2 B_{\mu}C^{\nu}v^{\mu\sigma}-2 B_{\mu}C^{\sigma}v^{\mu\nu}\pm 2 B_{\mu}C_{\lambda}\epsilon^{\mu\nu\sigma\lambda}-
\\
\nonumber
C^{\nu}C_{\lambda}w^{\sigma\lambda}+C^{\sigma}C_{\lambda}w^{\nu\lambda}-C^{\lambda}C_{\lambda}w^{\nu\sigma}+C_{\mu}C^{\nu}w^{\mu\sigma}-C_{\mu}C^{\sigma}w^{\mu\nu}=
\\
\nonumber
-2 B^{\nu}B_{\lambda}u^{\sigma\lambda}+2 B^{\sigma}B_{\lambda}u^{\nu\lambda}-B^{\lambda}B_{\lambda}u^{\nu\sigma}-2 i B^{\nu}C^{\sigma}+2 i B^{\sigma}C^{\nu}-
\\
\nonumber
2 B^{\nu}C_{\lambda}v^{\sigma\lambda}+2 B^{\sigma}C_{\lambda}v^{\nu\lambda}-2 B^{\lambda}C_{\lambda}v^{\nu\sigma}+2 B_{\mu}C^{\nu}v^{\mu\sigma}-2 B_{\mu}C^{\sigma}v^{\mu\nu}\pm 2 B_{\mu}C_{\lambda}\epsilon^{\mu\nu\sigma\lambda}-
\\
2 C^{\nu}C_{\lambda}w^{\sigma\lambda}+2 C^{\sigma}C_{\lambda}w^{\nu\lambda}-C^{\lambda}C_{\lambda}w^{\nu\sigma},
\end{eqnarray}
as, for example,
\begin{equation}\label{eq:dual44b}
-B^{\nu}B_{\lambda}u^{\sigma\lambda}+B_{\mu}B^{\nu}u^{\mu\sigma}=-B^{\nu}B_{\lambda}u^{\sigma\lambda}+B^{\nu}B_{\lambda}u^{\lambda\sigma}=-2 B^{\nu}B_{\lambda}u^{\sigma\lambda}.
\end{equation}
Thus, we can express $\psi_{\pm}^{\mu\nu}$ (and, finally, the current) via the scalar function  $\psi_{\mp}^{\mu\nu}u_{\mu\nu}$ (and its derivatives), and the results do not depend on the choice of the value of the square root in (\ref{eq:dual21b}), as, for example, products $B^{\nu}C_{\lambda}$  do not depend on the choice of the square root.

\section{\label{sec:level1n2}Formulation in terms of antisymmetric second-rank tensors}
Let us now formulate the tensor equivalent of the Dirac equation. First, we choose a constant non-zero antisymmetric second-rank tensor $u^{\mu\nu}$ satisfying the following conditions:
\begin{equation}\label{eq:dual45b}
\left(u^{\mu\nu}\right)=\left(\mp i\hodge u^{\mu\nu}\right),u^{\mu\nu}u_{\mu\nu}=0.
\end{equation}
Let us then denote the unknown  scalar function of the tensor equation $\varphi_u$ (instead of $\bar{\xi}\psi$). Then let us choose any constant antisymmetric second-rank tensor $v^{\mu\nu}$ satisfying the following conditions:
\begin{equation}\label{eq:dual46b}
\left(v^{\mu\nu}\right)=\left(\mp i\hodge v^{\mu\nu}\right),v^{\mu\nu}u_{\mu\nu}=0,u^{\mu}_{\;\:\sigma}v^{\sigma\nu}=- i u^{\mu\nu}.
\end{equation}
If we use again 3D vectors $\bm{u},\bm{v}, \bm{w}$, corresponding to the antisymmetric second-rank tensors $u^{\mu\nu},v^{\mu\nu},w^{\mu\nu}$, such that, for example, $\bm{u}=(u_1,u_2,u_3)=(u^{01},u^{02},u^{03})$, one can check that (\ref{eq:dual45b}) implies $(\bm{u}\cdot\bm{u})=0$ and (\ref{eq:dual46b}) implies $(\bm{v}\cdot\bm{u})=0$. One can also check that the last equality in (\ref{eq:dual46b}) implies
\begin{equation}\label{eq:dual46e2}
\bm{u}=\mp\bm{u}\times\bm{v}=\mp(u_2 v_3-u_3 v_2,-u_1 v_3+u_3 v_1,u_1 v_2-u_2 v_1),
\end{equation}
so
\begin{equation}\label{eq:dual46e}
\bm{u}=\pm\bm{v}\times\bm{u}=\bm{v}\times(\bm{v}\times\bm{u})=(\bm{v}\cdot\bm{u})\bm{v}-(\bm{v}\cdot\bm{v})\bm{u}=-(\bm{v}\cdot\bm{v})\bm{u},
\end{equation}
yielding $(\bm{v}\cdot\bm{v})=-1$ and $v^{\mu\nu}v_{\mu\nu}=4$. One can check that the last equality in (\ref{eq:dual46b}) is generally (if $u^{01}\neq 0$) equivalent to
\begin{equation}\label{eq:dual46c}
u^{03}v^{02}-u^{02}v^{03}= \pm u^{01}.
\end{equation}

Then, using the procedure from Section \ref{sec:level1a1}, we can calculate a constant antisymmetric second-rank tensor $w^{\mu\nu}$ satisfying the following conditions:
\begin{equation}\label{eq:dual47b}
\left(w^{\mu\nu}\right)=\left(\mp i\hodge w^{\mu\nu}\right),v^{\mu\nu}w_{\mu\nu}=0,w^{\mu\nu}w_{\mu\nu}=0,u^{\mu\nu}w_{\mu\nu}=-8.
\end{equation}  
Using (\ref{eq:sim2}) and (\ref{eq:dual14}), we obtain the following linear tensor equivalent of the Dirac equation:
\begin{equation}\label{eq:dual48b}
((2\Box'-F_{\mu\nu}v^{\mu\nu})(F_{\mu\nu}u^{\mu\nu})^{-1}(2\Box'
+F_{\mu\nu}v^{\mu\nu})+F_{\mu\nu}w^{\mu\nu})\varphi_u=0.
\end{equation}
The current can be restored using the procedure from Section \ref{sec:level1a1}. In particular, based on (\ref{eq:dual21b}), we should use the following:
\begin{equation}\label{eq:dual48c}
\varphi_u=\left(-\frac{\psi_\mp^{\mu\nu}u_{\mu\nu}}{8}\right)^{\frac{1}{2}}.
\end{equation}
Let us provide examples of the tensors $u^{\mu\nu},v^{\mu\nu},w^{\mu\nu}$, which play a role similar to that of $\gamma$-matrices in the spinor formulation of the Dirac equation. One of the choices for the tensors to be used with the upper signs in the equations like (\ref{eq:dual45b}) are:
\begin{eqnarray}\label{eq:dual48d}
\left(u^{\mu\nu}\right)=\left( \begin{array}{cccc}
0 & i & 1 & 0\\
-i & 0 & 0 & -i\\
-1 &0 & 0 & -1\\
0 &i & 1 &0
 \end{array} \right),
\left(v^{\mu\nu}\right)=\left( \begin{array}{cccc}
0 & 0 & 0 & -i\\
0& 0 & 1 & 0\\
0 &-1 & 0 & 0\\
i &0 & 0 &0
 \end{array} \right),
\left(w^{\mu\nu}\right)=\left( \begin{array}{cccc}
0 & -i & 1 & 0\\
i & 0 & 0 & -i\\
-1 &0 & 0 & 1\\
0 &i & -1 &0
 \end{array} \right).
\end{eqnarray}
One of the choices for the tensors to be used with the lower signs in the equations like (\ref{eq:dual45b}) are:
\begin{eqnarray}\label{eq:dual48e}
\left(u^{\mu\nu}\right)=\left( \begin{array}{cccc}
0 & i & 1 & 0\\
-i & 0 & 0 & i\\
-1 &0 & 0 & 1\\
0 &-i & -1 &0
 \end{array} \right),
\left(v^{\mu\nu}\right)=\left( \begin{array}{cccc}
0 & 0 & 0 & i\\
0& 0 & 1 & 0\\
0 &-1 & 0 & 0\\
-i &0 & 0 &0
 \end{array} \right),
\left(w^{\mu\nu}\right)=\left( \begin{array}{cccc}
0 & -i & 1 & 0\\
i & 0 & 0 & i\\
-1 &0 & 0 &-1\\
0 &-i & 1 &0
 \end{array} \right).
\end{eqnarray}
Let us prove that (\ref{eq:dual48b}) is equivalent to the Dirac equation. For example, if we use the upper signs in (\ref{eq:dual45b},\ref{eq:dual46b},\ref{eq:dual47b}),
\begin{eqnarray}\label{eq:dual49b}
\left(u^{\mu\nu}\right)=\left( \begin{array}{cccc}
0 & u^{01} & u^{02} & u^{03}\\
-u^{01} & 0 & i u^{03} & -i u^{02}\\
-u^{02} &-i u^{03} & 0 & i u^{01}\\
-u^{03} &i u^{02} & -i u^{01} &0
 \end{array} \right)=
\left( \begin{array}{cccc}
0 & u_1 & u_2 & u_3\\
-u_1 & 0 & i u_3 & -i u_2\\
-u_2 &-i u_3 & 0 & i u_1\\
-u_3 &i u_2 & -i u_1 &0
 \end{array} \right),
\end{eqnarray}
let us find such spinor 
\begin{equation}\label{eq:dual50b}
\xi=\left( \begin{array}{c}
\xi_1\\
\xi_2\\
0\\
0\end{array}\right)
\end{equation}
that
\begin{eqnarray}\label{eq:dual51b}
\nonumber
\left(u^{\mu\nu}\right)=\left(\bar{\xi}\sigma^{\mu\nu}\xi^c\right)=
\\
\left( \begin{array}{cccc}
0 & i((\xi_1^{*})^2-(\xi_2^{*})^2) & (\xi_1^{*})^2+(\xi_2^{*})^2 & -2 i\xi_1^{*}\xi_2^{*}\\
-i((\xi_1^{*})^2-(\xi_2^{*})^2) & 0 & 2 \xi_1^{*}\xi_2^{*} & -i ( (\xi_1^{*})^2+(\xi_2^{*})^2)\\
 -(\xi_1^{*})^2-(\xi_2^{*})^2 &-2 \xi_1^{*}\xi_2^{*} & 0 & -(\xi_1^{*})^2+(\xi_2^{*})^2)\\
2 i\xi_1^{*}\xi_2^{*} &i ( (\xi_1^{*})^2+(\xi_2^{*})^2) & (\xi_1^{*})^2-(\xi_2^{*})^2 &0
 \end{array} \right).
\end{eqnarray}
This is equivalent to the following system of equations for $\xi_1^*$ and $\xi_2^*$:
\begin{equation}\label{eq:dual52b}
u_1=i((\xi_1^{*})^2-(\xi_2^{*})^2),u_2=(\xi_1^{*})^2+(\xi_2^{*})^2,u_3=-2 i\xi_1^{*}\xi_2^{*}.
\end{equation}
As $(u_1)^2+(u_2)^2+(u_3)^2=0$, this system always has solutions and defines the spinor $\xi$ up to a factor $\pm 1$:
\begin{equation}\label{eq:dual53b}
\xi^*_1=\left(\frac{-i u_1+u_2}{2}\right)^\frac{1}{2},\xi^*_2=\frac{u_3}{(-2 i \xi^*_1)},
\end{equation}
or, if $-i u_1+u_2=0$,
\begin{equation}\label{eq:dual54b}
\xi^*_2=\left(\frac{i u_1+u^2}{2}\right)^\frac{1}{2},\xi^*_1=\frac{u_3}{(-2 i \xi^*_2)}.
\end{equation}
Let us choose one of the solutions and find such spinor
\begin{equation}\label{eq:dual55b}
\eta=\left( \begin{array}{c}
\eta_1\\
\eta_2\\
0\\
0\end{array}\right)
\end{equation}
that
\begin{eqnarray}\label{eq:dual56b}
\nonumber
\left(v^{\mu\nu}\right)=\left( \begin{array}{cccc}
0 & v^{01} & v^{02} & v^{03}\\
-v^{01} & 0 & i v^{03} & -i v^{02}\\
-v^{02} &-i v^{03} & 0 & i v^{01}\\
-v^{03} &i v^{02} & -i v^{01} &0
 \end{array} \right)=
\left( \begin{array}{cccc}
0 & v_1 & v_2 & v_3\\
-v_1 & 0 & i v_3 & -i v_2\\
-v_2 &-i v_3 & 0 & i v_1\\
-v_3 &i v_2 & -i v_1 &0
 \end{array} \right)=
\\
\left(\bar{\xi}\sigma^{\mu\nu}\eta^c\right)=
\left( \begin{array}{cccc}
0 & i(\eta_1^{*}\xi_1^{*}-\eta_2^{*}\xi_2^{*}) & \eta_1^{*}\xi_1^{*}+\eta_2^{*}\xi_2^{*} & -i(\eta_2^{*}\xi_1^{*}+\eta_1^{*}\xi_2^{*} )\\
- i(\eta_1^{*}\xi_1^{*}-\eta_2^{*}\xi_2^{*}) & 0 & \eta_2^{*}\xi_1^{*}+\eta_1^{*}\xi_2^{*}  & -i(\eta_1^{*}\xi_1^{*}+\eta_2^{*}\xi_2^{*})\\
 -\eta_1^{*}\xi_1^{*}-\eta_2^{*}\xi_2^{*} &-\eta_2^{*}\xi_1^{*}-\eta_1^{*}\xi_2^{*}  & 0 & -\eta_1^{*}\xi_1^{*}+\eta_2^{*}\xi_2^{*}\\
i(\eta_2^{*}\xi_1^{*}+\eta_1^{*}\xi_2^{*} )&i(\eta_1^{*}\xi_1^{*}+\eta_2^{*}\xi_2^{*}) & \eta_1^{*}\xi_1^{*}-\eta_2^{*}\xi_2^{*} &0
 \end{array} \right).
\end{eqnarray}

 This is equivalent to the following system of equations for $\eta_1^*$ and $\eta_2^*$:
\begin{equation}\label{eq:dual57b}
v_1=i(\eta_1^{*}\xi_1^{*}-\eta_2^{*}\xi_2^{*}) ,v_2=\eta_1^{*}\xi_1^{*}+\eta_2^{*}\xi_2^{*},v_3=-i(\eta_2^{*}\xi_1^{*}+\eta_1^{*}\xi_2^{*} ).
\end{equation}
One can check that, due to the properties of $\bm{u}$ and $\bm{v}$, these equations are compatible and have one solution for the spinor $\eta$:
\begin{equation}\label{eq:dual58b}
\eta_1^{*}=\frac{-i v_1+v_2}{2 \xi_1^{*}},\eta_2^{*}=\frac{i v_1+v_2}{2\xi_2^{*}},
\end{equation}
or, if $\xi_1^*=0$,
\begin{equation}\label{eq:dual59b}
\eta_1^{*}=\frac{v_3}{-i \xi_2^{*}},\eta_2^{*}=\frac{v_1}{-i\xi_2^{*}},
\end{equation}
or, if $\xi_2^*=0$,
\begin{equation}\label{eq:dual60b}
\eta_1^{*}=\frac{v_1}{i \xi_1^{*}},\eta_2^{*}=\frac{v_3}{-i\xi_1^{*}}.
\end{equation}
Then
\begin{eqnarray}\label{eq:dual61b}
\nonumber
\bar{\xi}\eta^c=\eta_2^*\xi_1^*-\eta_1^*\xi_2^*=\frac{i v_1+v_2}{2\xi_2^*}\xi_1^*-\frac{-i v_1+v_2}{2\xi_1^*}\xi_2^*=
\\
\nonumber
\frac{(i v_1+v_2)(\xi_1^*)^2-(-i v_1+v_2)(\xi_2^*)^2}{2\xi_1\xi_2^*}=
\frac{(i v_1+v_2)\frac{-i u_1+u_2}{2}-(-i v_1+v_2)\frac{(u_3)^2}{(-2 i)^2\frac{-i u_1+u_2}{2}}}{i u_3}=
\\
\nonumber
\frac{(i v_1+v_2)\frac{-i u_1+u_2}{2}-(-i v_1+v_2)\frac{-(u_1)^2-(u_2)^2}{-2(-i u_1+u_2)}}{i u_3}=
\\
\frac{(i v_1+v_2)(-i u_1+u_2)+(-i v_1+v_2)(-i u_1-u_2)}{2 i u_3}=\frac{v_1 u_2-v_2 u_1}{u_3}=1,
\end{eqnarray}
due to (\ref{eq:dual46e2}).

As was shown in Section \ref{sec:level1a1}, if antisymmetric tensors $u^{\mu\nu}$ and $v^{\mu\nu}$ satisfy (\ref{eq:dual17}) and  (\ref{eq:dual1}), and $u^{\mu\nu}\neq 0$, there is only one antisymmetric tensor $w^{\mu\nu}$ satisfying  (\ref{eq:dual16}) and  (\ref{eq:dual1}). On the other hand, the tensor $\bar{\xi}\sigma^{\mu\nu}\eta^c$ satisfies  (\ref{eq:dual16}) and  (\ref{eq:dual1}), so $\bar{\xi}\sigma^{\mu\nu}\eta^c=w^{\mu\nu}$. Using this and (\ref{eq:li4},\ref{eq:simp},\ref{eq:dual51b},\ref{eq:dual56b}), we obtain from (\ref{eq:dual48b}):
\begin{equation}\label{eq:dual62b}
((\Box'-f_{\xi\eta})f_{\xi\xi}^{-1}(\Box'
+f_{\xi\eta})+f_{\eta\eta})\varphi_u=0,
\end{equation}
which coincides with the equivalent (\ref{eq:sim2}) of the Dirac equation after we identify $\varphi_u$ with $\bar{\xi}\psi$.

\section{\label{sec:level1n3}Formulation in terms of 3D vectors}
Now let us rewrite the Dirac equation in terms of complex 3D vectors. For example, instead of the antisymmetric second-rank tensors $u^{\mu\nu},v^{\mu\nu},w^{\mu\nu}$, we are going to use 3D vectors, such as $\bm{u}=(u_1,u_2,u_3)$,where  $u_i=u^{0 i}$. Lorentz transformations correspond to 3D rotations of such vectors through complex angles.

We require
\begin{equation}\label{eq:dual63b}
(\bm{u}\cdot\bm{u})=0,(\bm{u}\cdot\bm{v})=0,\bm{u}=\mp\bm{u}\times\bm{v}.
\end{equation}
One can check that the last equality of (\ref{eq:dual63b}) generally yields
\begin{equation}\label{eq:dual64b}
u^{03}v^{02}-u^{02}v^{03}= \pm u^{01}
\end{equation}
and
\begin{equation}\label{eq:dual65b}
(\bm{v}\cdot\bm{v})=-1.
\end{equation}
We also require
\begin{equation}\label{eq:dual66b}
(\bm{u}\cdot\bm{w})=2,(\bm{v}\cdot\bm{w})=0,(\bm{w}\cdot\bm{w})=0.
\end{equation}
Adapting the calculation of $w^{\mu\nu}$ in  Section \ref{sec:level1a1}, we can obtain, using (\ref{eq:dual24a},\ref{eq:dual5},\ref{eq:dual6},\ref{eq:dual7},\ref{eq:dual8}):
\begin{equation}\label{eq:dual67b}
\bm{w}=-\frac{(\bm{v}\cdot\bm{u}^*)^2}{(\bm{u}\cdot\bm{u}^*)^2}\bm{u}+2\frac{(\bm{v}\cdot\bm{u}^*)}{(\bm{u}\cdot\bm{u}^*)}\bm{v}+\frac{2}{(\bm{u}\cdot\bm{u}^*)}\bm{u}^*,
\end{equation}
where $\bm{u}^*=(u_1^*,u_2^*,u_3^*)$.Then, using (\ref{eq:d8n1}), one can check that, e.g., 
\begin{equation}\label{eq:dual68b}
F_{\mu\nu}u^{\mu\nu}=2(\bm{u}\cdot(\bm{E}\mp i \bm{H})),
\end{equation}
where $\bm{E}=(E^1,E^2,E^3),\bm{H}=(H^1,H^2,H^3)$, introduce vector $\bm{F}=\bm{E}\mp i \bm{H}$, and use  (\ref{eq:dual48b}) to obtain
\begin{equation}\label{eq:dual69b}
((\Box'-(\bm{F}\cdot\bm{v}))(\bm{F}\cdot\bm{u})^{-1}(\Box'+(\bm{F}\cdot\bm{v}))+(\bm{F}\cdot\bm{w}))\varphi_u=0.
\end{equation} 
Again, the current can be restored using the procedure similar to that in Section \ref{sec:level1a1}. In particular, based on (\ref{eq:dual21b}), we should use the following:
\begin{equation}\label{eq:dual70b}
\varphi_u=\left(\frac{(\bm{\psi}_\mp\cdot\bm{u})}{2}\right)^{\frac{1}{2}},
\end{equation}
where vector $\bm{\psi}_\mp=(\psi_\mp^{0 1},\psi_\mp^{0 2},\psi_\mp^{0 3})$.

Let us provide examples of the vectors $\bm{u},\bm{v},\bm{w}$, which play a role similar to that of $\gamma$-matrices in the spinor formulation of the Dirac equation. One of the choices for the vectors are (cf. (\ref{eq:dual48d},\ref{eq:dual48e})):
\begin{equation}\label{eq:dual71b}
\bm{u}=(i,1,0),\bm{v}=(0,0,\mp i),\bm{w}=(-i,1,0).
\end{equation}
\maketitle
\section{\label{sec:level1c}Conclusion}

Previously ~\cite{Akhmeteli-JMP,Akhm2015,Akhmeteliqr}, the author showed that the Dirac equation is generally equivalent to a fourth-order linear equation for juct one component. On the other hand, chiral spinors can be represented by antisymmetric second-rank tensors ~\cite{Whit}. Thus, for a Dirac spinor $\psi$, one can build tensors $\psi^{\mu\nu}_{\pm}$ corresponding to chiral spinors $\psi_{\pm}$, such that $\psi=\psi_++\psi_-$ and $\gamma^5\psi_{\pm}=\pm\psi_{\pm}$. Then we introduce constant antisymmetric second-rank tensors $u^{\mu\nu},v^{\mu\nu}$ satisfying certain tensor conditions. The choice of $u^{\mu\nu}$ determines the component for which we obtain the linear equivalent of the Dirac equation, and different choices of $v^{\mu\nu}$ lead to equivalent equations. The choice of $u^{\mu\nu}$ and $v^{\mu\nu}$ determine the choice of another constant antisymmetric second-rank tensor $w^{\mu\nu}$  satisfying certain tensor conditions. Eventually, we derive a linear tensor equivalent of the Dirac equation for just one component
\begin{equation}\label{eq:dual72b}
\nonumber
\varphi_u=\left(-\frac{\psi_\mp^{\mu\nu}u_{\mu\nu}}{8}\right)^{\frac{1}{2}}.
\end{equation}
 There is a popular phrase: ``A spinor is a square root of a vector'' (~\cite{Wheeler}, Chapter II, para. 4). In our case, one can say that a square root of a scalar is a scalar.

Let us make some comments.
\begin{itemize}
    \item There are two slightly different versions of the formalism depending on whether we start with $\psi^{\mu\nu}_+$ or $\psi^{\mu\nu}_-$.
    \item When we have a solution of the linear tensor equivalent of the Dirac equation, there is an unequivocal recipe for calculation of the Dirac current.
    \item The linear tensor equivalent of the Dirac equation is an equation for one component, which component is ( in general) complex, but it can be made real by a gauge transformation (at least locally).
\end{itemize}
The author hopes that the linear tensor equivalent of the Dirac equation for one component will be useful for some applications and for better understanding of the Dirac equation's ``magic''.

\section*{Acknowledgments}

The author is grateful to A. Yu. Kamenshchik and A. D. Shatkus for their interest in this work and valuable remarks.

\end{document}